\documentclass[conference]{ieeeconf}

\IEEEoverridecommandlockouts   
\usepackage{setspace}
\usepackage{graphicx}
\usepackage{amssymb}
\usepackage{amsmath}
\usepackage{epstopdf}
\usepackage{dcolumn}
\usepackage{mathtools}
\usepackage{threeparttable}
\usepackage{standalone}
\usepackage{booktabs, dcolumn}
\usepackage{multirow}
\usepackage{algorithm}
 \usepackage{algpseudocode}
\usepackage{listings}
\usepackage{ragged2e}
\usepackage[usenames,dvipsnames]{color}
\usepackage{accents}
\usepackage{breqn}
\usepackage{float}
\usepackage{tikz,calc} 
\usepackage{amsfonts}
\usepackage{latexsym} 

\newtheorem{mydef}{Definition}
\usepackage{multicol}  

\definecolor{BlueLUH}{cmyk}{1.0,0.2,0,0}
\colorlet{LightBlue}{BlueLUH!20!white}

\usepackage{soul}

\usetikzlibrary{arrows,automata,calc}
\newcommand{\mat}[1]{\mathbf{#1}}
\newcommand{\pfr}[1]{\mathcal{#1}}

\usepackage[linkcolor=blue, colorlinks]{hyperref} 

\title{\LARGE \bf
Control Improvisation with Probabilistic Temporal Specifications 
}
\author{Ilge Akkaya, Daniel J. Fremont, Rafael Valle,\\ Alexandre Donz\'e, Edward A. Lee, and Sanjit A. Seshia% <-this % stops a space
\thanks{All authors are with the Department of Electrical Engineering and Computer Sciences, UC Berkeley, Berkeley, CA 94720, USA
        {\tt\small \{ilgea, dfremont, rafaelvalle, donze, eal, sseshia\}@berkeley.edu}}%
}

\begin{document}
\maketitle

\setstretch{1.075}
% \maketitle
% \thispagestyle{empty}
% \pagestyle{empty}

%%%%%%%%%%%%%%%%%%%%%%%%%%%%%%%%%%%%%%%%%%%%%%%%%%%%%%%%%%%%%%%%%%%%%%%%%%%%%%%%
\begin{abstract}
We consider the problem of generating randomized control sequences for
complex networked systems typically actuated by human agents. Our approach leverages
a concept known as control improvisation, which is based on
a combination of data-driven learning and controller synthesis from formal specifications. 
We learn from existing data a generative model (for instance, an explicit-duration hidden Markov model, or EDHMM)  and then supervise this model
in order to guarantee that the generated sequences satisfy some
desirable specifications given in Probabilistic Computation Tree Logic
(PCTL). We present an implementation of our approach and apply
it to the problem of mimicking the use of lighting appliances in a
residential unit, with potential applications to home security
and resource management. We present experimental results showing
that our approach produces realistic control sequences, similar to 
recorded data based on human actuation, while satisfying suitable formal requirements.
%We validate the resulting stochastic model by
%estimating the divergence of its distribution from that of the real
%system and by model checking it against the PCTL specifications using
%the tool PRISM. 
\end{abstract}


%==========================================================================
\section{Introduction}
\label{sec:intro}

The promise of the emerging Internet of Things (IoT) is to leverage
the programmability of connected devices to enable applications
such as connected smart vehicles,
occupancy-based automated HVAC control, 
autonomous robotic surveillance, and much more.
However, this promise cannot be realized without better tools for
the automated programming and control of a network of devices ---
computational platforms, sensors, and actuators. 
Traditionally, this problem has been approached from two different
angles.
The first approach is to be {\em data-driven}, leveraging the ability
of devices and sensors to collect vast amounts of data about their
operation and environments, and using learning algorithms to adjust
the working of these devices to optimize desired objectives. This
approach, exemplified by devices such as smart learning thermostats,
can be very effective in many settings, but typically cannot give any
guarantees of correct operation.
The second approach is to be {\em model-driven}, where accurate models
of the devices and their operating environment are used to define a
suitable control problem. A controller is then synthesized to
guarantee correct operation under specified environment conditions.
However, such an approach is difficult in settings where such accurate
models are hard to come by. Moreover, strong correctness guarantees may not
be needed in all cases.

%However, in many IoT settings, the primary concerns differ from the 
%traditional case. 
Consider, for instance, the application domain of home automation.
More specifically, consider a scenario where one is designing the controller for
a home security system that controls the lighting (and possibly other appliances)
in a home when the occupants are away. One might want to program the
system to mimic typical human behavior in terms of turning lights on
and off. As human behavior is somewhat random, varying day to day, one might want
the controller to exhibit random behavior. However, completely random
control may be undesirable, since the system must obey certain
time-of-day behavioral patterns, and respect correlations between devices. For these requirements, a data-driven approach where one learns a
randomized controller mimicking human behavior seems like a good
fit. It is important to note, though, that such an application may
also have constraints for which provable guarantees are needed, 
such as limits on energy consumption being obeyed with high
probability, or that multiple appliances never be turned on simultaneously.
A model-based approach is desirable for these.
Thus, the overall need is to {\em blend data and models} to synthesize
a control strategy that obeys certain {\em hard constraints} (that must
always be satisfied), certain {\em soft constraints} (that must be ``mostly
satisfied'') and certain {\em randomness requirements} on system behavior.

This setting has important differences from typical control problems.
For example, in traditional supervisory control, the goal is typically
to synthesize a control strategy ensuring that certain
mathematically-specified (formal) requirements hold on the entity
being controlled (the ``plant'').  
Moreover, the generated sequence of control inputs is typically
completely determined by the 
state of the plant. Predictability and correctness guarantees are
important concerns. 
In contrast, in the home automation application sketched above,
predictability is not that important. Indeed, the system's behavior
must be random, within constraints.
Moreover, the source of randomness (behavior of human occupants)
differs from home to home, and so this cannot be pre-programmed.

This form of randomized control is suitable
for human-in-the-loop systems or applications where randomness is desirable for
reasons of security, privacy, or diversity. Application domains other
than the home automation setting described above 
include microgrid scheduling
\cite{strelec2012modeling,gamarra2015computational} and robotic art
\cite{assayag2004using}. 
In the former, randomness can provide
some diversity of load behavior, hence making the grid more efficient
in terms of peak power shaving and more resilient  
to correlated faults or attacks on deterministic behavior. 
For the latter case, there is growing interest in augmenting 
human performances with computational aids, such as in automatic
synthesis and improvisation of music~\cite{donze2014machine}. 
All these applications share the property of requiring some randomness
while maintaining behavior within specified constraints.
Additionally, the human-in-the-loop applications can benefit from 
data-driven methods.
Streams of time-stamped data from devices can be used to learn 
semantic models capturing behavioral correlations amongst them for
further use in programming and control.

In this paper, we show how a recently-proposed formalism termed {\em
control improvisation}~\cite{fremont_et_al:LIPIcs:2015:5659} can be suitably
adapted to address the problem of randomized control for IoT
systems. We consider the specific setting of a system whose components
can be controlled either by humans or automatically. Human control of
devices generates data comprising streams of time-stamped events.
From such data, we show how one can learn a nominal randomized
controller respecting certain constraints present in the
data including correlations between behavior of interacting
components. We also show how additional constraints can be enforced on
the output of the controller using temporal logic specifications and
verification methods based on model
checking~\cite{Baier-2008book,KNP11}. 
We apply our approach to the problem of randomized control of home
appliances described above. We present simulated experimental results 
for the problem of lighting control based on data from the UK
Domestic Appliance-Level Electricity (UK-DALE) dataset \cite{UK-DALE}.
Our approach produces realistic control sequences, similar to 
recorded data based on human actuation, while also satisfying suitable
formal requirements. 

% \shorten{The rest of the paper is organized as follows: in Section}~\ref{sec:backgnd}, we give some preliminary background on the formalisms used throughout the paper.
% We formally define our problem as an instance of control improvisation, and describe our technique for solving it, in Section~\ref{sec:ci-pctl}.
% Next, Section~\ref{sec:results} presents detailed experimental results from a case study on lighting improvisation.
% Finally, we conclude the paper with related work and future directions in Sections~\ref{sec:relwork} and \ref{sec:conclusion} respectively.

%==========================================================================
\section{Background}
\label{sec:backgnd}
We introduce relevant background material that the present paper builds upon
and establish notation for use in the rest of the paper.

\subsection{Discrete-Event Systems with Hidden States}

%\ilge{(ADD DISCRETE-EVENT AND/OR HMM CITATIONS?)}

Our work focuses on control of systems whose behavior can be described by a sequence of timestamped \emph{events}.
An event $e$ is a tuple $\langle\tau,v\rangle\in T\times V$, where $T$ is a totally ordered set of time stamps and $V$ is a finite set of values.
We define a {\em signal} to be a set of events, where $T$ imposes an ordering relation on the events occurring within the signal \cite{LeeSan:98:TaggedSignal}.
 
We define the \emph{state} of such a system to take values from a finite set of distinct states, where {\em events} are emitted by state transitions. In many systems, the underlying events
and states are hidden, and all that can be observed is some
function of the state. We term this the {\em observation}. This function can be time-dependent
and probabilistic, so that a single state can produce many
different observations. We assume that the number of possible
observation values is finite (this can be enforced in continuous
systems by discretization), and that observations are made at discrete time steps. A sequence of observations over
time generated by a behavior of the system is called a \emph{trace}.

\begin{figure}\centering
\includegraphics[width=\columnwidth,clip]{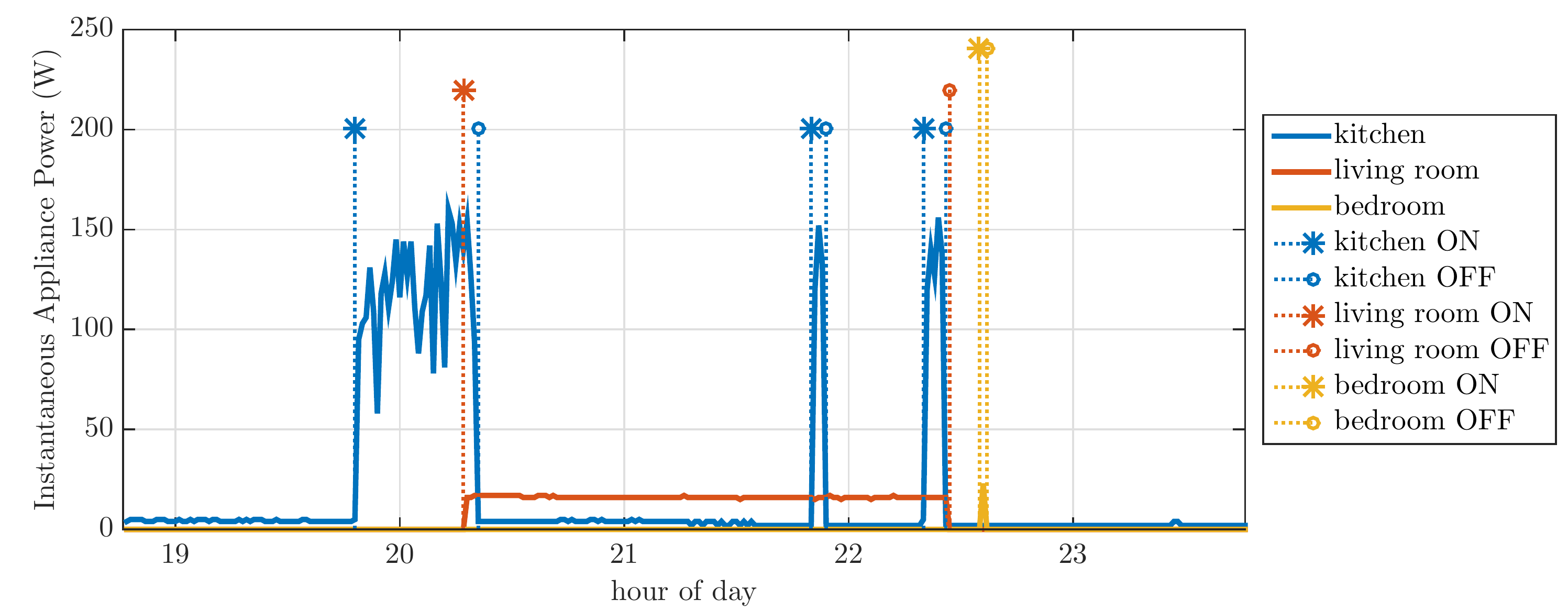}
\caption{Sample appliance power trace}
\label{fig:applianceTrace}
\end{figure}

An example of such a trace that captures the power consumption of three appliances is given in Figure~\ref{fig:applianceTrace}.
The events related to each appliance, which can either be an ``ON'' or an ``OFF'' event in this case, are annotated on the sub-traces.
Each state change of the system triggers an event.
Consider, for example, that the hidden state in this scenario captures the current status of a set of physical appliances and that all appliances are initially turned off.
The kitchen appliance being turned on at 19:50 pm causes an ``ON'' event to be emitted, and triggers a state change in the system, where in the new state, the kitchen appliance is on, and the other two appliances are off.
The system stays in this state until any appliance triggers a state transition.
In such a scenario, it may be that the only information available from the system are traces of the instantaneous appliance power consumptions.
Given these traces, one can {\em infer} the state of the system and which events may have happened at particular times.

% Our work focuses on control of systems whose behavior can be described by a sequence of time stamped \emph{events}. An event $e$ is a tuple $<\tau,v>\in T\times V$, where $T$ is a totally ordered set of time stamps and $V$ is a finite set of values. We define a {\em signal} to be a set of events, where $T$ imposes an ordering relation on the events occurring within the signal \cite{lee1998framework}.
% \ilge{should we mention the tagged signal model and the above paragraph? the state definition here is ambiguous. if we instead limit the scope to periodic signals where all time steps are an integer multiple of a sampling period, $T \subset \mathbb{Z}$, then it's simpler, because the state will then be defined at each integer multiple of the period, events can only happen at these times, and state transitions will be triggered by events.} 
%An event begins at a particular time, lasts for its \emph{duration}, and then ends: for example, ``the kitchen lights being on from 6 to 7 PM'' could be an event.
% We treat the case where there are a finite number of possible events, all occurring at times which are multiples of some underlying time step.
%We define the \emph{state} of such a system to be a function of events that have happened so far, and state transitions to be triggered by events. 
%In many systems, the underlying events and states are \emph{hidden}, and all that can be observed is some function of the state.  

% \Ra{This paper is based on synthesizing PFAs, but never once is
% a PFA defined or even described in plain English. One
% cannot expect an audience to be familiar with this term. }
\subsection{Control Improvisation}
\label{sec:control-improv}
% \Ra{ 

% The definition of a control improvisation problem is poorly
% written and the conditions (a)-(c) are poorly motivated. 
% It is also not entirely clear that control improvisation is
% a necessary or sufficient framework to solve the problem of
% developing an automated appliance controller for security
% applications. In particular, the control improvisation
% framework imposes a minimum amount of variability in the
% traces that are produced, but it is not clear that this
% requirement meshes particularly well with the data.  What
% if the data itself demonstrates very little variability?
% Then it seems that the switch from human-based control to
% the improviser would be noticed.  In general, this work
% would be stronger if there was some sort of condition on
% statistical indistinguishability as a hard constraint.

% It is difficult to understand from what is written what is
% meant by``comput[ing] the values of missing parameters". 
% Please explain this further before the experiment section}
%\subsection{Formalization as Control Improvisation}
% \Rc{- From my understanding, control improvisation is the
% strategy of generating randomized control sequences to make
% a system satisfy some given specifications. Maybe the
% author should clearly define what control improvisation is,
% since it is new to the readers and appears so many times in
% the paper. }
The {\em control improvisation problem}, defined formally
in~\cite{fremont_et_al:LIPIcs:2015:5659}, can be seen as the problem of generating
a random sequence of control actions subject to {\em hard} and {\em
soft} constraints, while satisfying a {\em randomness requirement}.
The hard constraints may, for example, encode
safety requirements on the system that must always be obeyed. The soft
constraints can encode requirements that may occasionally be
violated. The randomness requirement ensures that no control sequence
occurs with too high probability.
% --- in other words, the output of
%the controller ``exhibits variability.''

This problem is a natural fit to the applications of interest in this paper, as our
end goal is to randomize the control of discrete-event systems subject to both
constraints enforcing the presence of certain learned behaviors (hard
constraints), and probabilistic requirements upper bounding the observations (soft constraints).
In the lighting control scenario we consider later, for example, we effectively learn a hard constraint preventing multiple appliances from being toggled at exactly the same time, since this never occurs in the training data.
We also use soft constraints to limit the probability that the hourly power consumption exceeds desired bounds.

More formally, the control improvisation problem is defined as
follows. This is generalized from the definition in~\cite{fremont_et_al:LIPIcs:2015:5659} to allow multiple soft constraints with different probabilities.
\begin{mydef} \label{def:control-improv}
An instance of the control improvisation (CI) problem is composed of 
(i) a language $I$ of \emph{improvisations} that are strings over a
finite alphabet $\Sigma$, i.e., $I \subseteq \Sigma^*$, and
(ii) finitely many subsets $\mathcal{A}_i \subseteq I$ for $i \in \{1, \dots, n\}$.
These sets can be presented, for example, as finite automata, but for our purposes in this paper the details are unimportant (see \cite{fremont_et_al:LIPIcs:2015:5659} for a thorough discussion).

Given error probability bounds ${\boldsymbol \epsilon} = (\epsilon_1, \dots, \epsilon_n)$ with  $\epsilon_i \in [0,1]$,
and a probability bound $\rho \in (0,1]$, a distribution $D: \Sigma^*
\rightarrow [0,1]$ with support set $S$ is an
\emph{$(\boldsymbol{\epsilon},\rho)$-improvising distribution} if  
\begin{align*}
  (a)~&S \subseteq I & \text{(hard constraints)},\\
  (b)~&\forall w \in S, ~D(w) \leq \rho & \text{(randomness)},\\
  (c)~&\forall i, P[w\in \mathcal{A}_i \; | \; w \leftarrow D] \geq 1-\epsilon_i & \text{(soft constraints)},
\end{align*}
\noindent where $w \leftarrow D$ indicates that $w$ is drawn from the distribution $D$.
An \emph{$(\boldsymbol{\epsilon},\rho)$-improviser}, or simply an \emph{improviser}, is a probabilistic algorithm
generating strings in $\Sigma^*$ whose
output distribution is an $(\boldsymbol{\epsilon},\rho)$-improvising distribution.
For example, this algorithm could be a
Markov chain generating random strings in $\Sigma^*$.
The control improvisation problem is, given the tuple
$(I,\{\mathcal{A}_i\},\boldsymbol{\epsilon},\rho)$, 
to generate such an improviser.

\end{mydef}
\subsection{Explicit-Duration Hidden Markov Models}
\label{sec:edhmm}

\begin{figure}
\centering
\resizebox{0.85\columnwidth}{!}{
\begin{tikzpicture}[->,>=stealth',shorten >=1pt,auto,node distance=1.8cm,inner sep=0pt, minimum size=0pt,
                    semithick]
  \tikzstyle{every state}=[fill=none,draw=black,text=black]
  \tikzstyle{every observedstate}=[fill=grey,draw=black,text=black]
  %\node[state,draw=none] (d0)                    {$\pi_d$};  
  \node[state]    (d1) {$d_1$}; 
  \node[state,draw=none]    (d2) [ right of=d1] {\dots};
  \node[state]    (dt) [ right of=d2] {$d_{t-1}$};
  \node[state]    (dtp1) [ right of=dt] {$d_{t}$};
  \node[state,draw=none]    (dtp2) [ right of=dtp1] {\dots};
  \node[state]    (dT) [ right of=dtp2] {$d_T$};

  %\node[state,draw=none]   (x0) [ below of=d0] {$\pi_x$};  
  \node[state]    (x1) [ below of=d1] {$x_1$}; 
  \node[state,draw=none]    (x2) [ right of=x1] {\dots};
  \node[state]    (xt) [ right of=x2] {$x_{t-1}$};
  \node[state]    (xtp1) [ right of=xt] {$x_{t}$};
  \node[state,draw=none]    (xtp2) [ right of=xtp1] {\dots};
  \node[state]    (xT) [ right of=xtp2] {$x_T$};
  \node[state,fill=gray!40]   (y1) [ below of=x1] {$y_1$}; 
  \node[state,draw=none]    (y2) [ right of=y1] {\dots};
  \node[state,fill=gray!40]   (yt) [ right of=y2] {$y_{t-1}$};
  \node[state,fill=gray!40]   (ytp1) [ right of=yt] {$y_{t}$};
  \node[state,draw=none]    (ytp2) [ right of=ytp1] {\dots};
  \node[state,fill=gray!40]   (yT) [ right of=ytp2] {$y_T$};
  \node[] (priord) [left of=d1,node distance=1cm] {$\pi_d$};
  \node[] (priorx) [left of=x1,node distance=1cm] {$\pi_x$};

  \path 
      (d1) edge              node {} (d2)
      (d1) edge          node {} (x2)
      (d2) edge              node {} (dt)
      (dt) edge              node {} (dtp1)
      (dtp1) edge              node {} (dtp2)
      (dtp2) edge              node {} (dT)
    %(x0) edge node {} (x1)
    (x1) edge node {} (x2)  
    (x2) edge              node {} (xt)
      (xt) edge              node {} (xtp1)
      (xtp1) edge              node {} (xtp2)
      (xtp2) edge              node {} (xT)
      (x1) edge node {} (d1)
      (x1) edge node {} (y1)
      (d2) edge node {} (xt)
      (xt) edge node {} (yt)
      (xt) edge node {} (dt)
      (dt) edge node {} (xtp1)
      (xtp1) edge node {} (dtp1)
      (xtp1) edge node {} (ytp1)
      (dtp1) edge node {} (xtp2)
      (dtp2) edge node {} (xT)
      (xT) edge node {} (yT)
      (xT) edge node {} (dT);
\end{tikzpicture}}
\caption{Graphical model representation of an EDHMM}
\label{edhmm_simple}
\end{figure}
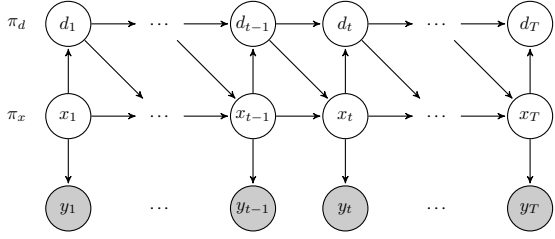

%% Original text:
% In data-driven controller synthesis, it is essential that the
% learning model captures relevant properties of the underlying
% data. In applications where the data exhibits an underlying
% but hidden state space, and observations can be assumed to
% be drawn from densities depending only on the current state,
% a Hidden Markov Model (HMM) is a good fit.

%A useful state-space model of observation streams generated by a dynamical system assumes an underlying hidden state, which evolves as a function of input {\em events} and time.  Hidden Markov Models (HMM) are a widely-used instance of these models, which assume a discrete hidden state-space that evolves according to a Markov process, where observed data is modeled to have a state-dependent distribution \cite{rabiner1989tutorial}. 

{
In data-driven controller synthesis, it is essential that the
learning model captures relevant properties of the underlying system based on observed data. 
For probabilistic inference in dynamical systems whose state is only observable via state-dependent data,  Hidden Markov Models (HMM) have been a widely used tool. An HMM is characterized by a hidden state variable subject to Markov dynamics, observable via state-dependent noisy observations. 
However, in many applications, including those of interest to this
paper, the Markov
assumption on the hidden state space is insufficient, since part of
the underlying problem structure lies in the \emph{durations} of 
events. In such scenarios, the model quality
can be improved significantly by the use of semi-Markov models.} In
this study, we will specifically consider the Explicit-Duration Hidden
Markov Model (EDHMM) \cite{rabiner1989tutorial}. 
These models are an extension of
HMMs that, in addition to modeling the hidden state
space as a Markov chain, also introduces the duration spent within each state as another hidden variable of the Bayesian network. The graphical model representation of a general EDHMM is shown in Figure~\ref{edhmm_simple}.

% \Ra{Please motivate why durations should belong to a finite
% set.  Why are they not 
% continuous? Is this explicitly a sampled-data system?}
The standard definition of an EDHMM models hidden state and its duration to be discrete hidden variables.
The state dependent observations can be drawn from either a discrete or continuous distribution, often referred to as an {\em emission} distribution.
In this paper, we assume that the possible observations are quantized as necessary so that the emission distributions are discrete.

An EDHMM with discrete emissions observed for $T$ time steps is characterized by a partially observed set of variables $(\mathbf{x},\mathbf{d},\mathbf{y})$ = $(x_1,\dots,x_T$, $d_1,\dots,d_T$, $y_1,\dots,y_T)$.
Each $x_i$ indicates the hidden state of the model at time $i$ from a finite state space $\mathcal{X}$, which for notational convenience we assume to be the set $\{1, \dots, N\}$.
The value $d_i \in \{1, \dots, D\}$ denotes the remaining duration in the hidden state, where $D$ is the maximum possible state duration.
Finally, each $y_i$ is an observation drawn from a discrete alphabet $\Sigma=\{v_1,\dots,v_M\}$.
The joint probability distribution imposed by the EDHMM over these variables can be written as
\vspace{-0.1cm}

\begin{align*}
P(\mathbf{x},\mathbf{d},\mathbf{y}) &= \resizebox{.75\hsize}{!}{$p(x_1)p(d_1)\displaystyle\prod_{t=2}^{T}p(x_t|x_{t-1},d_{t-1})p(d_t|d_{t-1},x_t) \prod_{t=1}^T{p(y_t|x_t})$} \nonumber \\
&= \resizebox{.75\hsize}{!}{$\pi_{x}\pi_{d}\displaystyle\prod_{t=2}^{T}p(x_t|x_{t-1},d_{t-1})p(d_t|d_{t-1},x_t) \prod_{t=1}^T{p(y_t|x_t)}$},
\end{align*}

\noindent where $p(x_1)\triangleq \pi_x$ %= \{p(q_1=1),...,p(q_1=N)\}$ 
and $p(d_1)\triangleq \pi_d$ %=\{p(d_1)=1,...,p(d_1)=D\}$ 
are the priors on the hidden state and duration distributions, respectively. The conditional state and duration dynamics are given by
% \Ra{
% The notation used to define the prior distributions doesn't
% make sense to me.  The prior is a set?  Shouldn't it be a
% function?}
\begin{align}
%p(q_1) &\sim \mu_{q_0} \quad~ p(d_1) \sim \mu_{d_0}\\
p(x_t|x_{t-1},d_{t-1}) &\triangleq \begin{cases} p(x_t| x_{t-1}) &\text{if}~d_{t-1} = 1 \\ \delta(x_t,x_{t-1}) & otherwise   \end{cases} \label{pqs}\\
p(d_t|d_{t-1},x_t) &\triangleq \begin{cases} p(d_t|x_t) &\mbox{if }  d_{t-1} = 1 \\ \delta({d_t,d_{t-1}-1}) & otherwise\end{cases} \,,\label{pds}
\end{align}

\noindent where $\delta(\cdot,\cdot)$ is the Kronecker delta function.
Equations (\ref{pqs}) and (\ref{pds}) specify the current state $x_t$ and the remaining duration $d_t$ for that state as a function of the previous state and its remaining duration.
Unless the remaining duration at the previous state is equal to 1, the state will remain unchanged across time steps, while at each step within the state, the remaining duration is decremented by 1.
When the remaining duration is 1, the next state is sampled from a transition probability distribution $p(x_t|x_{t-1})$, while the remaining duration at $x_t$ is sampled from a state-dependent duration distribution $p(d_t|x_t)$.
All self-transition probabilities are set to zero: $p(x_t|x_{t-1})=0$ if $x_t=x_{t-1}$.
For compactness of notation, for all $x_t,x_{t-1} \in \{1,\dots,N\}$, $d_t \in \{1,\dots,D\}$, and $y_t\in\{v_1,\dots,v_M\}$ we define probabilities $a_{x_{t-1}, x_t}$, $b_{x_t, y_t}$, and $c_{x_t, d_t}$ so that
% \Rc{- It took me a while to understand (1) and (2), which
% basically say that if d_{t-1} is larger than 1, the state
% will keep the same value while the duration will decrease
% by 1; otherwise, the model will evolve according to some
% given transition probability. Maybe I am slow but I would
% recommend the authors to add some explanation to help the
% readers to understand the equations and some others if
% necessary. }
\begin{align*}
p(x_t|x_{t-1}) &= \begin{cases}a_{x_{t-1},x_{t}} &\text{if}~x_t\neq x_{t-1} \\ 0 &\text{otherwise}\,\end{cases},\\
p(y_t|x_t) &= b_{x_t,y_t}\,, \\
p(d_t|x_t) &= c_{x_t,d_t}\,.
\end{align*}
We consolidate these probabilities into an $N \times N$ state transition matrix $A \triangleq (a_{ij})$, an $N \times M$ emission probability matrix $B \triangleq (b_{ij})$, and an $N \times D$ duration probability matrix $C \triangleq (c_{ij})$.

% \noindent where $a_{q_{t-1},q_{t}} \triangleq P(q_t$ is the entry at row $i=q_{t-1}$ and column $j=q_{t}$ of the {state \emph{transition probability matrix}} $A$, and $\delta(\cdot)$ is the Kronecker delta function. For the following formulations, we assume $A$ is a homogeneous matrix ($ A_{t} = A ~\forall t\in\{1,...,T\}$), however, the non-homogeneous case also follows the derivation with minor modifications. Furthermore, note that, since self-transitions are captured by $p(d_t|d_{t-1},q_t)$, $a_{ii} = 0, ~ \forall i \in \{1,...,N\}$ where $N$ is the number of hidden states in the model definition.
% \Ra{Overall, the definition of the EDHMM is quite confusing.  I
% think that many readers would be unfamiliar with the term
% emission probability.  Also, I think that while the
% definitions of the conditional probabilities given here
% make sense, it takes too much time to parse what they mean.
%  A simple example or plain English explanation would make
% this section significantly more readable.}

% The graphical model representation of an EDHMM is given by Figure
% \ref{edhmm_sup}. 

The procedure to obtain the EDHMM parameter set $\lambda~=~\{\pi_x,\pi_d,A,B,C\}$, given the observed sequence $\boldsymbol y$, is often referred to as the \emph{parameter estimation} problem, which in the general Bayesian inference setting seeks to assign the parameters of a model so that it best explains given training data.
More precisely, given a trace $(y_1,\dots,y_T)$, parameter estimation approximates the optimal parameter set $\lambda^*$ such that
\begin{align*}
\lambda^* &= \text{arg}~\underset{\lambda}{\text{max}}~p(y_1, \dots, y_T \; | \; \lambda)\,. 
\end{align*}
This procedure can be extended to estimate parameters from multiple traces, provided that the traces are aligned so that the first observation of each trace corresponds to the same initial state.
This ensures that the state prior will be correctly captured \cite{rabiner1989tutorial}.
% \red{ILGE: what about training from multiple traces?}
% \ilge{Training on multiple traces doesn't change the underlying algorithm, however, some implementation details arise. For example, the traces need to begin at the same time step, because we will be estimating a single prior to determine the beginning state. They can, however, be of different lengths, the algorithm will still work}
In the case of the EDHMM, parameter estimation can be done with a variant of the well-known Expectation-Maximization (EM) algorithm for HMM. The detailed formulation is presented in \cite{yu2010hidden}. 

% \red{ILGE: do we need the following line?}
% The parameter estimation problem is a means to learn
% an automaton structure given observed data, and therefore,
% it becomes one way of performing data-driven automaton
% synthesis.

\subsubsection{EDHMM with Non-homogeneous Hidden Dynamics}

The general definition of an EDHMM is useful in modeling hidden state dynamics encoded with explicit duration information.
However, in many applications where the state dynamics model behaviors that exhibit seasonality, it can be useful to train separate state transition and duration distributions for different time intervals.
As an example, we consider the case where the dynamics exhibit a dependence on the hour of the day, so that for each hour $h \in \{1, \dots, 24\}$ we have different probability matrices $A_h$ and $C_h$. 

Estimating the parameters of an EDHMM with hourly dynamics requires an additional input sequence $\{h_1,\dots,h_T\}$, where each $h_i \in \{1,...,24\}$ labels at which hour of the day the observation $y_i$ was collected.
Given the observation and hour label streams, training follows the same EM-based estimation procedure as in \cite{yu2010hidden}, with the difference that parameters $A_h$ and $C_h$ are estimated using the training data subsequences collected within hour $h$. 
%The rest of the parameter set is trained on the complete training sequence.

% \red{ILGE: do we need this last paragraph? I simplified the}
% \red{notation in Section III.F so that $a_{i,j}^h$ isn't needed any more}
% \ilge{But if we don't define these as a function of hour and refer to these as $a_{i,j}$ alone, it is not clear that these probabilities depend on the hour in general. The notation in III.F is fine but should we not at least say that we have 24 of each of these matrices for clarity?}
The EDHMM with hourly dynamics will be given by a parameter set $\lambda=\{\pi_x,\pi_d,\{A_h\},B,\{C_h\}\}$, where $\{A_h\}$ and $\{C_h\}$ are the transition and duration distribution matrices valid for hour $h\in\{1,...,24\}$ such that

\begin{align*}
a_{i,j}^l &\triangleq p(x_t=j|x_{t-1}=i,d_{t-1}=1,h_{t-1}=l) \\
c_{i,d}^l &\triangleq p(d_t=d|x_t =i,d_{t-1}=1,h_{t}=l)
\end{align*}

\noindent where $A_l=(a_{ij}^l)$ and $C_l=(c_{i,d}^l)$ are the hourly transition and duration probability matrices for hour $l$.

\subsection{Probabilistic Model Checking}

Our approach relies on the use of a verification method known as
probabilistic model checking, which determines if a probabilistic
model (such as a Markov chain or Markov decision process) satisfies a
formal specification expressed in a probabilistic temporal logic.
We give here a high-level overview of the relevant concepts for this
paper. The reader is referred to the book by Baier and
Katoen~\cite{Baier-2008book} for further details.
% \Ra{
% The paragraph describing PCTL semantics is also confusing. 
% Formal logics are as yet not a widely used tool in control
% theory, so this section should be friendlier.  Although
% conjunction and negation are standard in most propositional
% logics, a short definition of what these
% symbols mean would greatly help readers who are not used to
%  such formalisms.  A short example of a PCTL formula with a
% plain English explanation would also be helpful.}
For our application, we employ probabilistic computation tree logic
(PCTL). The syntax of this logic is as follows:
\begin{align}
\phi & ::=True \mid \omega \mid \neg\phi \mid \phi_{1}\wedge\phi_{2} \mid P_{\Join p}\left[\psi\right] & \mbox{state formulas}\notag\\
\psi & ::= \mat{\pfr{X}}\phi \mid \phi_{1}\;\pfr{U}^{\leq k}\phi_{2} \mid \phi_{1}\;\pfr{U}\phi_{2} & \mbox{path formulas}\notag
\end{align}
\noindent where $\omega\in\Omega$ is an atomic proposition,
$\Join\in\{\leq, <, \geq, >\}$, $p\in [0,1]$ and $k\in \mathbb{N}$.
State formulas are interpreted at states of a probabilistic model; if
not specified explicitly, this is assumed to be the initial state of
the model.
Path formulas $\psi$ use the \emph{Next} $\left(\mat{\pfr{X}}\right)$,
\emph{Bounded Until} $\left(\pfr{U}^{\leq k}\right)$ and
\emph{Unbounded Until} $\left(\pfr{U}\right)$ operators.  
These formulas are evaluated over computations (paths)
and are only allowed as parameters to the $P_{\Join p}\left[\psi\right]$ operator. 
%The size $\mathcal{Q}$ of a PCTL formula is defined as the number of Boolean connectives plus
%the number of temporal operators in the formula. 
%For the \emph{Bounded Until} operator, we denote separately the maximum time bound that 
%appears in the formula as $k_{max}$.
Additional temporal operators, $\pfr{G}$, denoting ``globally'', and
$\pfr{F}$ denoting ``finally'', are defined as follows: $\pfr{F} \phi
\triangleq True \, \pfr{U} \, \phi$ and $\pfr{G} \phi \triangleq \neg \, \pfr{F} \, \neg \phi$.

We describe the semantics informally; the formal details are available
in~\cite{Baier-2008book}. A path formula of the form $\mat{\pfr{X}}\phi$ holds on
a path if state formula $\phi$ holds on the second state of that path.
A path formula of the form $\phi_{1}\;\pfr{U}^{\leq k}\phi_{2}$ holds on a path
if the state formula $\phi_2$ holds eventually at some state on that path within $k$ steps 
of the first state, with
$\phi_1$ holding at every preceding state. 
The semantics of $\phi_{1}\;\pfr{U}\phi_{2}$ is similar without the ``within $k$ steps'' requirement.
The semantics of state formulas is standard for all propositional formulas. The only
case worth elaborating involves the probabilistic operator: $P_{\Join p}\left[\psi\right]$ holds 
at a state $s$ if the probability $q$ that path formula $\psi$ holds for any execution beginning at $s$
satisfies the relation $q \Join p$.

A probabilistic model checker, such as PRISM~\cite{KNP11}, can check whether a probabilistic
model satisfies a specification
in PCTL. Moreover, it can also compute the probability that a temporal logic formula holds in a model, as well as synthesize missing model parameters so as to satisfy a specification. 
We show in Sec.~\ref{encoding-prism} how an EDHMM can be encoded as 
a Markov chain and thereby as a suitable input model to PRISM.

%==========================================================================

%==========================================================================
% \section{Learning a Probabilistic Finite Automaton from Data}
% \label{sec:pfa-learning}

%z\input{pfa-learning}

%==========================================================================
\section{Control Improvisation with Probabilistic Temporal Specifications}
\label{sec:ci-pctl}

Now we define the problem tackled in this
paper, and describe the approach we take to solve it.

\subsection{Problem Definition and Solution Approach} 

We begin with a set of traces of a discrete-event system whose set of events is known, but whose dynamics are not.
Our goal is to randomly generate new traces with similar characteristics to the given ones.
Furthermore, we want to be able to enforce two kinds of constraints:
\begin{itemize}
 \item \emph{Hard} constraints that the traces must always satisfy, forbidding transitions between states that never occur in the input traces.
 For example, if no part of the input traces can be explained as a particular state transition $t$, then we want to assume that $t$ is impossible and not generate any string that is only possible using it.
 
 \item \emph{Soft} constraints that need only be satisfied with some given probability.
 We focus on systems whose observations are \emph{costs}, for example power consumption, and assume soft constraints which put upper bounds on the cost at a particular time, or accumulated over a time period.
\end{itemize}
In the next section, we will formalize this problem as an instance of control improvisation. 
First, however, we summarize our solution approach, which consists of three main steps:
\begin{enumerate}
\item {\em Data-Driven Modeling:} From the given traces, learn an EDHMM
representing the time-dependent dynamics of the underlying system.
%The language of all output strings generated by these automata form
%$I$, the hard constraint.
The EDHMM effectively applies hard constraints on our generation procedure by eliminating all strings assigned zero probability.

\item {\em Probabilistic Model Checking:} 
Using a probabilistic model checker, we compute the probability that a behavior of the candidate improviser obtained in the previous step will satisfy the soft constraints. 
If this is sufficiently high, we return the EDHMM as our generative model.

 \item {\em Scenario-Based Model Calibration:}
 Otherwise, we apply heuristics that increase the probability by modifying the EDHMM parameters, and return to step (2).
% In the case when model checking concludes that the soft constraints are not satisfied with sufficiently high probability, we apply heuristics to increase the probability by modifying the EDHMM parameters.
% These are applied incrementally until the model checker proves that the soft constraints are satisfied with at least the desired probability.

\end{enumerate}

% \begin{figure*}[!htdb]
% \centering
% \includegraphics[width=0.8\textwidth,clip,trim=0cm 2cm 0cm 0cm]{workflow2}
% \caption{Algorithmic workflow}
% \label{fig:workflow}
% \end{figure*}

\begin{figure}
\centering
\resizebox{\columnwidth}{!}{
\begin{tikzpicture}[%
    auto, 
    block/.style={
      rectangle,
      draw=blue,
      thick,
      fill=blue!20,
      text width=3em,
      align=center,
      rounded corners,
      minimum height=2em
    },
    bigBlock/.style={
      rectangle,
      draw=black,
      thick,
      fill=blue!20,
      text width=3em,
      align=center,
      rounded corners,
      minimum height=1.5cm, 
      node distance = 3.5cm
    },
    block1/.style={
      rectangle,
      draw=blue,
      thick,
      fill=blue!20,
      text width=em,
      align=center,
      rounded corners,
      minimum height=2em
    },
    line/.style={
      draw,thick,
      -latex',
      shorten >=2pt
    } 
  ]

    \node[block,fill=yellow!20] (A1) {};
    \node[block,below of=A1] (A2) {};
    \node[block, fill=green!20, below of=A2] (A3) {};
    \node[fill=none,above of=A1] (title) {Time-Series Data}; 

    \node (rect) at (0,-1) [draw,thick,dashed,minimum width=2cm,minimum height=3.4cm,rounded corners] {};

    \node[bigBlock, text width=3cm,fill=none, right of=A2] (B) {EDHMM \\Parameter\\ Estimation};
    \node[bigBlock, text width=2.4cm,fill=none, right of=B] (C) {Candidate Improviser};
    \node[bigBlock, text width=2.4cm,node distance = 4.5cm,fill=none, right of=C] (D) {Probabilistic Model Checking};
    \node[bigBlock, fill=LightBlue,line width=0.8mm, text width=2.4cm, below of=D, node distance = 2.5cm] (E) {\large Control Improviser};

    \node[bigBlock, text width=4cm, node distance=2.6cm,fill=none, minimum height=1cm,below of=B] (F) {PCTL Properties};
    \node[right of=F, node distance = 5cm] (G) {};

    \draw[->,>=stealth,line width=1.5](rect.east) -- (B.west);
    \draw[->,>=stealth,line width=1.5](B.east) -- (C.west) ;

    \draw[->,>=stealth,line width=1.5](C.east) -- node[text width=1.8cm] { \small{~PRISM}\\~Code\\~Generation} (D.west) ;

    \draw[->,>=stealth,line width=1.5](D.south) -- (E.north) ;
    \draw[->,>=stealth,line width=1.5, bend left](D.north) to [out=290,in=250] node[above] {Scenario-based Model Calibration} (C.north); 
    \draw[->,>=stealth,line width=1.5](rect.south) |- node[left] {\Large $\Delta_{max}^h$} (F.west) ;
    \draw[->,>=stealth,line width=1.5](G.east) |- ($(D.south west)!0.25!(D.north west)$); 
    \draw[>=stealth,line width=1.5](F.east) -- (G.east); 
  \end{tikzpicture}}
  \caption{Algorithmic Workflow} \label{fig:workflow}
  \end{figure}
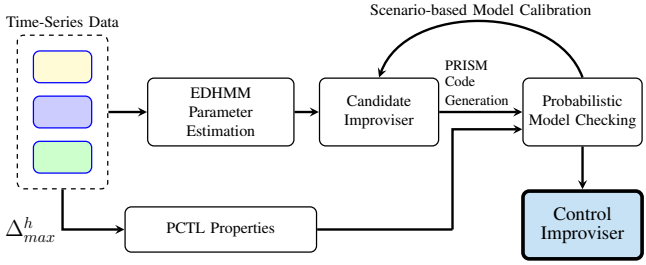

A high level algorithmic workflow is given by Figure~\ref{fig:workflow}. We elaborate on each of the steps in subsequent sections. 

\subsection{Formalization as a Control Improvisation Problem}

We can formalize the intuitive description above as an instance of the control improvisation (CI) problem described in Section \ref{sec:control-improv}.
To do so, we need to specify the alphabet $\Sigma$, languages $\mathcal{I}$ and $\mathcal{A}_i$, and parameters $\epsilon_i$ and $\rho$ that make up a CI instance.
\begin{description}
\item[$\Sigma$] Since we are learning from and want to generate traces, which are sequences of observations, we let $\Sigma$ be the set of all observations (i.e. those occurring anywhere in the input traces). \vspace{0.5\baselineskip}

\item[$I$] We let $\mathcal{I}$ consist of all traces that are assigned nonzero probability by the EDHMM\footnote{The definition of the CI problem given in \cite{fremont_et_al:LIPIcs:2015:5659} requires that $I$ be described by a finite automaton.
It is straightforward to build a nondeterministic finite automaton that accepts precisely those strings assigned nonzero probability by the EDHMM, but we will not describe the construction here since it is not needed for the technique used in this paper.}.
Since the CI problem requires any improviser to output only strings in $I$, this will ensure the hard constraints are always satisfied. \vspace{0.5\baselineskip}

\item[$\mathcal{A}_i, \epsilon_i$] We let $\mathcal{A}_i$ consist of all traces that satisfy the $i$-th soft constraint.
For instance in the lighting example, $\mathcal{A}_i$ could only contain  traces whose total power consumption within hour $i \in \{1, \dots, 24\}$ of the day never exceeds a given bound.
Then in the CI problem, $\epsilon_i$ is the greatest probability we are willing to tolerate of the improviser generating a trace violating the bound. \vspace{0.5\baselineskip}

\item[$\rho$] We can ensure that many different traces can be generated, and that no trace is generated too frequently, by picking a small value for $\rho$: the CI problem requires that no improvisation be generated with probability greater than $\rho$, and so that at least $1/\rho$ improvisations can be generated.
\end{description}

This CI problem captures the informal requirements we described earlier.
Now we need to show that our generation procedure is actually an improviser solving this problem according to the three conditions given in Definition \ref{def:control-improv}.
We consider each in turn.

 \subsubsection{Hard Constraints}

By definition, any string that we generate has nonzero probability according to the EDHMM and so is in $\mathcal{I}$.

 \subsubsection{Randomness Requirement}

As long as the EDHMM is ergodic (when converted to an ordinary Markov chain; see Section \ref{encoding-prism}), the probability of generating any particular string $w \in \Sigma^*$ goes to zero as its length goes to infinity.
So for any $\rho \in (0,1]$, we can satisfy the randomness requirement by generating sufficiently long strings.
We can efficiently detect when the EDHMM is not ergodic using standard graph algorithms, but this is unlikely to be necessary in practice for applications as lighting control.

 \subsubsection{Soft Constraints}\label{sec:soft-constraints}

Our procedure checks whether this requirement is satisfied using probabilistic model checking.
This requires encoding the sets $\mathcal{A}_i$ as PCTL formulas, and the EDHMM as a Markov chain (explained in Sections \ref{soft-constraints} and \ref{encoding-prism} respectively).
Once this has been done, the model checker computes the probability that a string generated by the EDHMM will be in $\mathcal{A}_i$.
If this probability is at least $1 - \epsilon_i$, then the EDHMM satisfies the soft constraint, and if this is true for each $i$, it is a valid improviser.
Otherwise, our procedure applies heuristics to modify the EDHMM, detailed in Section~\ref{sec:heurstics}.
As shown in that section, the heuristics decrease the expected accumulated cost, so that after sufficiently many applications the EDHMM will satisfy the soft constraints\footnote{Obviously, some soft constraints cannot be satisfied, for example one requiring that the cost at the first time step be less than the smallest possible cost of any state. See Section~\ref{sec:heurstics} for a precise statement.}.

Therefore, our generation procedure yields a valid improviser solving the CI problem we defined above.
We note that our technique has some further useful properties not captured by the CI problem.
In particular, we can easily disable particular transitions between hidden states by setting their probabilities to zero and normalizing remaining transition probabilities appropriately.
This could be useful, for example, when controlling an IoT system with unreliable components: if a component drops off the network or becomes otherwise unusable, we can disable all transitions to states in which that component is active.

%$\mathcal{L}(\mathcal{W}_I)$ denotes the language of $\mathcal{W}_I$. Let $A \subseteq I$ denote the set of \emph{admissible} improvisations, such that

\subsection{Learning an EDHMM from Traces}\label{sec:edhmmlearning}

The first step in our procedure is to learn an EDHMM from the input traces.
Since as explained in Section~\ref{sec:edhmm} we use an EDHMM with different transition matrices for each hour, every input trace $\{\mathbf{y}_1,\mathbf{y}_2,\dots,\mathbf{y}_T\}$ is augmented with a corresponding stream of labels $\{\tau_1,\dots,\tau_T\}$ indicating the hour of the day each observation was recorded.
Note that the observations need not be scalar costs, but could be vectors: for example, in our lighting experiments each observation was a $K$-tuple $\mathbf{y}_i = [y_{i,1}, \dots, y_{i,K}]^T$ containing instantaneous power readings from each of $K$ different appliances.

Given this training data, we perform EDHMM parameter estimation as described in Section~\ref{sec:edhmm}.
This yields a parameter set $\lambda=(\{A_h\},\{C_h\},B,\pi)$ where the matrices $\{A_h\}$ and $\{C_h\}$ give state transition and duration probabilities respectively for each hour $h \in \{1,\dots,24\}$.
The distribution of observations for each state is given by $B$, and $\pi$ is the prior on the state space.
%\red{ILGE: do we need to explain where the prior comes from?}
In this work we use categorical distributions for $B$ and $\{C_h\}$, although in other applications it may be appropriate to use parametric distributions.

Note that the parameter estimation process based on the EM algorithm is an iterative method; thus obtaining a reasonable parameter set depends on model convergence, which in turn requires sufficient training data.
In the case of an EDHMM with hourly transition matrices, if few events happen at certain hours it may not be possible to estimate some of the state transition and duration probabilities for those hours.
Many application-specific heuristics exist for handling such scenarios, as outlined in \cite{rabiner1989tutorial}.
The particular technique we used in our experiments is detailed in Section~\ref{sec:experimental-setup}.

\subsection{Encoding Soft Constraints as PCTL Formulas} \label{soft-constraints}

As mentioned earlier, we consider soft constraints which put upper bounds on the cost observed at a particular time or accumulated over a time period.
We illustrate how to encode upper bounds on the hourly cost --- other time periods are handled analogously.

Recall that our traces take the form $\{\mathbf{y}_1,\mathbf{y}_2,\dots,\mathbf{y}_T\}$ where each $\mathbf{y}_i$ is an observation, generally a vector $[y_{i,1},\dots,y_{i,K}]^{\text{T}}$ of costs.
Define $Y_i \triangleq \sum_{k=1}^K y_{i,k}$, the total cost at time step $i$.
Considering that the data is sampled at the rate of $N_s$ samples per hour, the total hourly cost accumulated up to time step $t$ is
$$\Delta = \sum_{N_s (\lceil t / N_s \rceil - 1) + 1 \le i \le t} Y_i .$$
In the next section, we show how a simple monitor added to the encoding of the EDHMM can maintain the value $\Delta$.

In order to be able to impose a different upper bound $\Delta_{max}^h$ on $\Delta$ for each hour $h$ of the day, we need to compute the current hour of the day as a function of the time step:
$$h(t) = mod(\lceil t / N_s \rceil - 1, 24)+1\,,$$
\noindent which holds if $t=1$ corresponds to the time step of the first sample collected within hour 1. 
Then we can write the soft constraint for hour $h$ as the following PCTL formula:
\begin{align}P_{\ge 1-\epsilon_h} \; \pfr{G} \left[ (h(t)=h) \Rightarrow (\Delta \leq \Delta_{max}^h) \right] .
\label{eq:pctl-hourly}
\end{align}
This simply asserts that with probability at least $1-\epsilon_h$, at every time step during hour $h$ the corresponding upper bound on $\Delta$ holds.
In practice we can omit the quantifier $P_{\ge 1-\epsilon_h}$ and ask the probabilistic model checker to compute the probability that the rest of the formula holds, instead of having to specify a particular $\epsilon_h$ ahead of time. 

%Given the predicate definitions, it is of interest to synthesize an improviser that satisfies the hourly PCTL properties
%\begin{equation}
%P_{\geq}(1-\epsilon^h)[\pfr{G} (e_t \leq {\mathcal{P}}_{max}^h)],\quad h=1,...,24
%\label{eqn:pctl}
%\end{equation}
%\noindent where $\epsilon^h \in [0,1]$ is an error probability upper bound of violating the path formula on word segments produced at hour h. Note that such improviser $\mathcal{W}_I$ generates a distribution $D:\Sigma^* \rightarrow [0,1]$, which is a $(\epsilon,\rho)$-improvising distribution, such that $\epsilon$ is the error probability upper bound of not satisfying (\ref{eqn:predConj}).

% The motivation behind defining hourly predicates for the lighting improvisation problem is that the EDHMM with hourly transition and duration distributions gives us an incremental way of controlling the likelihood of events happening at particular hours, which could be used to precisely control properties on hourly power consumption.  

% $$ \epsilon \triangleq \max_{h\in\{1,...,24\}}\epsilon_{max}^h\,.$$

\subsection{Encoding the EDHMM as a Markov Chain}\label{encoding-prism}

In this section, we discuss how the EDHMM can be represented as a Markov chain, so that the soft constraints can be verified using probabilistic model checking.

Ignoring the soft constraints for now, the interpretation of the EDHMM as a Markov chain follows the outline in Section~\ref{sec:edhmm}: we expand the state space with a new state variable $d \in \{1, \dots, D\}$ which keeps track of the remaining duration in the current hidden state $x \in \{1, \dots, N\}$.
When $d > 1$, we stay in $x$ for another time step, decrementing $d$.
Only when $d = 1$ do we transition to a new hidden state, picking the new value of $d$ from the corresponding duration distribution.

Since we use an extension of the EDHMM where state transition and duration probabilities depend on the current hour, we need to expand the state space further to keep track of time.
The state variable $t \in \{0, \dots, T\}$ indicates the current time step, with $t = 0$ being an initialization step in which the state is sampled from a state prior $\pi$.
Note that the domain of $t$ need not grow unboundedly with $T$: in our example where we use different transition probabilities for each of the 24 hours, we only need to track the time within a single day.

Finally, in order to detect when the soft constraints are violated, we need to monitor the total hourly cost $\Delta$ defined in the previous section.
We add the state variable $\Delta \in \{0, \dots, \Delta_{max} + 1\}$, where $\Delta_{max}$ is the largest of the hourly upper bounds $\Delta_{max}^h$ imposed by the soft constraints.
This range of values for $\Delta$ is clearly sufficient to detect when the total cost exceeds any of these bounds.
Maintaining the correct value of $\Delta$ is simple: at each time step we increase it by a cost sampled from the appropriate emission distribution, except when a new hour is starting, in which case we first reset it to zero.

Putting this all together, we obtain a Markov chain whose states are 4-tuples $(x,d,t,\Delta)$ with the state variables as described above.
The initial state is $(0,1,0,0)$.
Given the current state, the next state $(x',d',t',\Delta')$ is determined as follows:

% \red{ILGE: writing $d' \sim p(d'|x')=C_{h(t)}(x')$ looks confusing}
% \red{would it be okay to just write $d' \sim C_{h(t)}(x')$ ?}

\texttt{EDHMM:}
\begingroup
\addtolength{\jot}{-0.08cm}
 \begin{align*} 
% condition  
(t=0) &\rightarrow& x'\sim\pi_x & \;\wedge\\
&&d'\sim C_{h(t)}(x') & \;\wedge\\
%&& y_i' \sim p(y_i|x')=B(x')&\wedge\\
&&t'=t+1 & \\ 
(t>0)\wedge(d>1) &\rightarrow & x'=x &\;\wedge\\
&&d'=d-1 &\;\wedge\\
%&& y_i' \sim p(y_i|x')&\wedge \\
&& t'=t+1 & \\
(t>0)\wedge(d=1) &\rightarrow& x'\sim A_{h(t)}(x) &\;\wedge\\
&& d'\sim C_{h(t)}(x')&\;\wedge\\
%&& y_i' \sim p(y_i|x')&\wedge \\
&& t'=t+1 &
 \end{align*}

\endgroup

%\texttt{State Cost:}
% \begin{align*} 
% (t>0)&\rightarrow&y_i \sim p(y_i|x)=B(x)
% \end{align*}

\texttt{Cost Monitor:}
\begingroup
\addtolength{\jot}{-0.08cm}
 \begin{align*}  
(t=0) &\rightarrow&\Delta'=0\\
(t>0)\wedge(h(t')=h(t)) &\rightarrow& \Delta' = \Delta + \sum_{i=1}^K p_i,~\mathbf{p}\sim B(x)\\
(t>0)\wedge(h(t')\neq h(t))&\rightarrow&\Delta' =\sum_{i=1}^K p_i ,~\mathbf{p} \sim B(x)\,,
\end{align*}
\endgroup

\noindent where $h(t) = mod(\lceil t / N_s \rceil - 1, 24)+1$.

\subsection{Scenario-Based Model Calibration}\label{sec:heurstics}

The procedure described so far provides a way to obtain a generative model that captures the probabilistic nature of events and their duration characteristics in a physical system, and to verify that the model satisfies desired soft constraints.
However, the model may not satisfy these constraints with sufficiently high probability, particularly if the constraints are not always satisfied by the training data.
In terms of control improvisation, the error probability of our improviser for some soft constraint $i$ is greater than the desired $\epsilon_i$.
We now describe two general heuristics for calibrating the EDHMM to decrease the error probability while preserving the faithfulness of the improviser to the original data.
In particular, these heuristics do not introduce new behaviors: any trace that can be generated by the calibrated improviser could already be generated before calibration.
Since the soft constraints we consider place upper bounds on the observed costs, both heuristics seek to decrease the costs of some behavior of the improviser.

\subsubsection{Duration Calibration}
The duration distributions of the trained EDHMM, $\{C_h\}$, assume a maximum state duration $D$ that is enforced during the training process.
One simple way to decrease cost is to further restrict the duration distributions by truncating them beyond some threshold for some or all states.
An effective strategy in practice is to eliminate outliers in the duration distributions of states with high expected cost.

This heuristic has the advantage of leaving the transition probabilities of the model completely unchanged, and so is a relatively minor modification.
On the other hand, it cannot reduce the duration of a state below 1 time step.
So although it can eliminate some high-cost behaviors from the model, it is not guaranteed to eventually yield an improviser satisfying the soft constraints.

\subsubsection{Transition Calibration}

A different approach is to modify the state transition probabilities, making the model less likely to transition to a high cost state during certain hours of the day.
Specifically, we can limit the probability of transitioning from any state $i$ to a particular state $x_r$ during hour $h_r$ to be at most some value $p_r^i$.
We shift the removed probability mass to the transition leading to the state $x_{min}$ with least expected cost, which we assume is strictly less than that of $x_r$.
Writing the original transition probability matrix $A_{h_r}$ as $(a_{ij})$, we replace it in the EDHMM with a new matrix $\tilde{A}_{h_r} = (\tilde{a}_{ij})$ defined by

\begin{align*}
\tilde{a}_{ij} = 
\begin{cases} 
\min(p_r^i,a_{ij}) & \text{if}~j=x_r \\
a_{ij}+(a_{ix_r}-\min(p_r^i,a_{ij})) & \text{if}~ j=x_{min} \\
a_{ij} & \text{otherwise.}
\end{cases}\,
\end{align*}

\noindent Note that the second case ensures that the transition probabilities from any state $i \in \mathcal{X}$ are properly normalized.
Provided that the limits $p_r^i$ are chosen such that $\tilde{a}_{ix_r}< {a}_{ix_r}$ for some $i \in \mathcal{X}$, the heuristic will decrease the expected cost of a behavior generated by the improviser.

Applying the heuristic iteratively for every choice of $x_r \ne x_{min}$ and hour $h_r \in \{1,\dots,24\}$ will eventually result in an improviser that remains at the $x_{min}$ state for all time steps (assuming $x_{min}$ is the starting state).
Thus for any soft constraints which are true for behaviors that only stay at $x_{min}$, our procedure will eventually terminate and yield a valid improviser.
This over-simplified improviser is unlikely to model the original data well, but it is only attained as the limit of this heuristic: in practice, judicious choices of the state $x_r$ and limits $p_r^i$ can improve the error probability significantly in a few iterations without drastically changing the model.

%==========================================================================

\section{Experimental Results and \\Analysis}
\label{sec:results}

\subsection{Experimental Setup}
\label{sec:experimental-setup}

%Inputs to the CI Problem:

To demonstrate the control improvisation approach we have
described in Sec.~\ref{sec:ci-pctl}, we use the UK Domestic Appliance-Level Electricity
(UK-DALE) dataset \cite{UK-DALE}, which contains disaggregated time
series data representing instantaneous power consumptions of
residential appliances from 5 homes over a period of 3 years.  

We consider a lighting improvisation scenario over the three most-used
lighting appliances in a single residence, each from a separate room
of the house. The data is presented as a vector-valued power
consumption sequence $\mathbf y$ with a corresponding sequence of time
stamps $\boldsymbol \tau$.  
The input stream $\mathbf{y} = \{\mathbf{y}_1,\mathbf{y}_2,
\dots,\mathbf{y}_T\}$ consists of 3-tuples  

$$ {\bf y}_i = \begin{bmatrix} y_{i,1} \\ y_{i,2} \\ y_{i,3} \end{bmatrix} , \; i = 1, \dots,T \,,$$

\noindent where the values $y_{i,1}$, $y_{i,2}$, and $y_{i,3}$ are
instantaneous power readings with time stamp $\tau_i$ from the main
kitchen light, a dimmable living room light, and the bedroom light
respectively. 
The power readings were sampled with a period of 1 minute and are
measured in watts. 

In our experiments, we synthesized three improvisers from this data: one using an unmodified EDHMM, and two that were calibrated using the different kinds of heuristics described in Section \ref{sec:heurstics} to enforce soft constraints on hourly power consumption.
Below, we describe the specific choices that were made when implementing each of the three main steps of our procedure.

% \subsubsection{Algorithmic Workflow}
% ~\\ 

\subsubsection{Data-Driven Modeling}
We assume there are three sources of hidden events, corresponding to each of the three appliances being turned on or off.
This yields a hidden state space $\mathcal{X}$ with 8 states, one for each combination of active appliances.
Based on inspection of the dataset, we chose the maximum state duration to be 720 time steps (12 hours, sufficient to allow long periods when all appliances are off).
Since we used disaggregated data, our observations are 3-tuples of power consumptions (quantized to integer values as part of the dataset), which we assume fall in the alphabet $\Sigma = \Sigma_1 \times \Sigma_2 \times \Sigma_3$ where $\Sigma_1 = \{0, 1, \dots, 350\}$, $\Sigma_2 = \{0,1,\dots,20\}$, and $\Sigma_3 = \{0,1,\dots,30\}$ (the maximum consumptions for each appliance were again obtained by inspecting the dataset).
Having fixed these parameters (summarized in Table \ref{table:exp}), an EDHMM was trained from a 100-day subset of the data from one residence.
Several portions of this training data (for one appliance) are shown at the top of Figure \ref{fig:temporal}.

Note that for the specific case of lighting improvisation, 
since the power emission distributions of each appliance are independent,
%have an independent probability distribution from that of other appliances, 
$B\triangleq p({\bf y_t}|x_t)$, the learned emission probability matrix over vector-valued observations, can be written as
%\vspace{-0.2cm}
$$ B = p({\bf y_t}|x_t) = \prod_{k=1}^K p({y_{t,k}}|x_t)\,.$$

It should also be noted that following the training process, some of the state transition probabilities $\{A_h\}$ may remain unlearned, i.e., we may have 
$$\sum_{j=1}^{N}a_{i,j}^h =0$$ 
\noindent for some state $i\in\{1,...,N\}$.
This can occur, for example, when no state transitions from state $i$ happen during the hour $h$ in any of the input traces.
%\red{ILGE: see my email about the completion strategy}
Since it is key to capture the observed appliance behavior, we treat these incomplete distributions that are unobserved in the training data as behaviors that should also be absent from the set of improvised behaviors.
Consequently, we use a completion strategy that forces transitions to the state $x_{min}$ with the least expected cost (i.e. the state with all appliances off) in this scenario:
\begin{align*}
&\forall a_{i,j}^h\,, ~\text{where} \sum_{j=1}^{N}a_{i,j}^h =0, \\
 i,j &\in \{1,...,N\},~h \in \{1,...,24\}, \\
\tilde{a}_{i,j}^h &= \begin{cases} 1 & \text{if} ~j=x_{min}  \\ 0 & \text{otherwise}\end{cases}
\end{align*}
where $\tilde{a}_{i,j}^h$ is the adjusted state transition probability of switching from state $i$ to $j$ in hour $h$.
Note that in this case study, such incomplete parameter estimates arose only for early morning hours in which few state transitions were recorded (typically hours $h\in\{1,\dots,5\}$).
Having completed the transition probability matrices in this way, we obtain a fully specified EDHMM.

% \begin{table}
% \begin{tabular}{lllll}
% State Label & State Index & \multicolumn{3}{c}{Appliance State (0=OFF, 1=ON)}\\
% &&Kitchen & Living Room & Bedroom \\
% OFF & 1 & 0 & 0 & 0 \\
% B& 2 & 0 & 0 & 1 \\
% L& 3 & 0 & 1 & 0 \\
% BL& 4 & 0 & 1 & 1 \\
% K& 5 & 1 & 0 & 0 \\
% KB& 6 & 1 & 0 & 1 \\
% KL& 7 & 1 & 1 & 0 \\
% KLB& 8 & 1 & 1 & 1 \\
% \end{tabular}
% \caption{The Hidden State Enumerations for the Lighting EDHMM}
% \end{table}

\begin{table}
\centering \footnotesize
\begin{tabular}{ll}
\hline
{\bf Parameter ID} & {\bf Value} \\
\hline 
\hline
Data Source & UK DALE Dataset \\
\hline
House ID & house\_1 \\
\hline
\multirow{3}{*}{Appliance IDs} & kitchen\_lights \\
& livingroom\_s\_lamp \\
& bedroom\_ds\_lamp \\
\hline
Training Duration & 100 days \\
Training Start Date &  30 Jul 2013 19:07:56 GMT \\ 
Sampling Period ($T_s$) & 60 s \\
Training Sequence Length ($T$) & 144000 \\ 
\hline
Maximum State Duration ($D$) & $720$ \\
Appliance 1 Costs ($\Sigma_1$) & $\{0,1,\dots,350\}$\\ 
Appliance 2 Costs ($\Sigma_2$) & $\{0,1,\dots,20\}$\\ 
Appliance 3 Costs ($\Sigma_3$) & $\{0,1,\dots,30\}$\\ 
\hline
\multirow{8}{*}{State Labels} & OFF: All appliances off \\
& K: Kitchen on \\
& L: Living room on \\
& B: Bedroom on \\
& KL: Kitchen and living room on \\
& KB: Kitchen and bedroom on \\
& LB: Living room and bedroom on \\
& KLB: All appliances on \\
\hline
\end{tabular}
\caption{Parameters of the training dataset for EDHMM learning}\label{table:exp}
\end{table}

\begin{figure}[t]
\centering
\includegraphics[width=0.9\columnwidth]{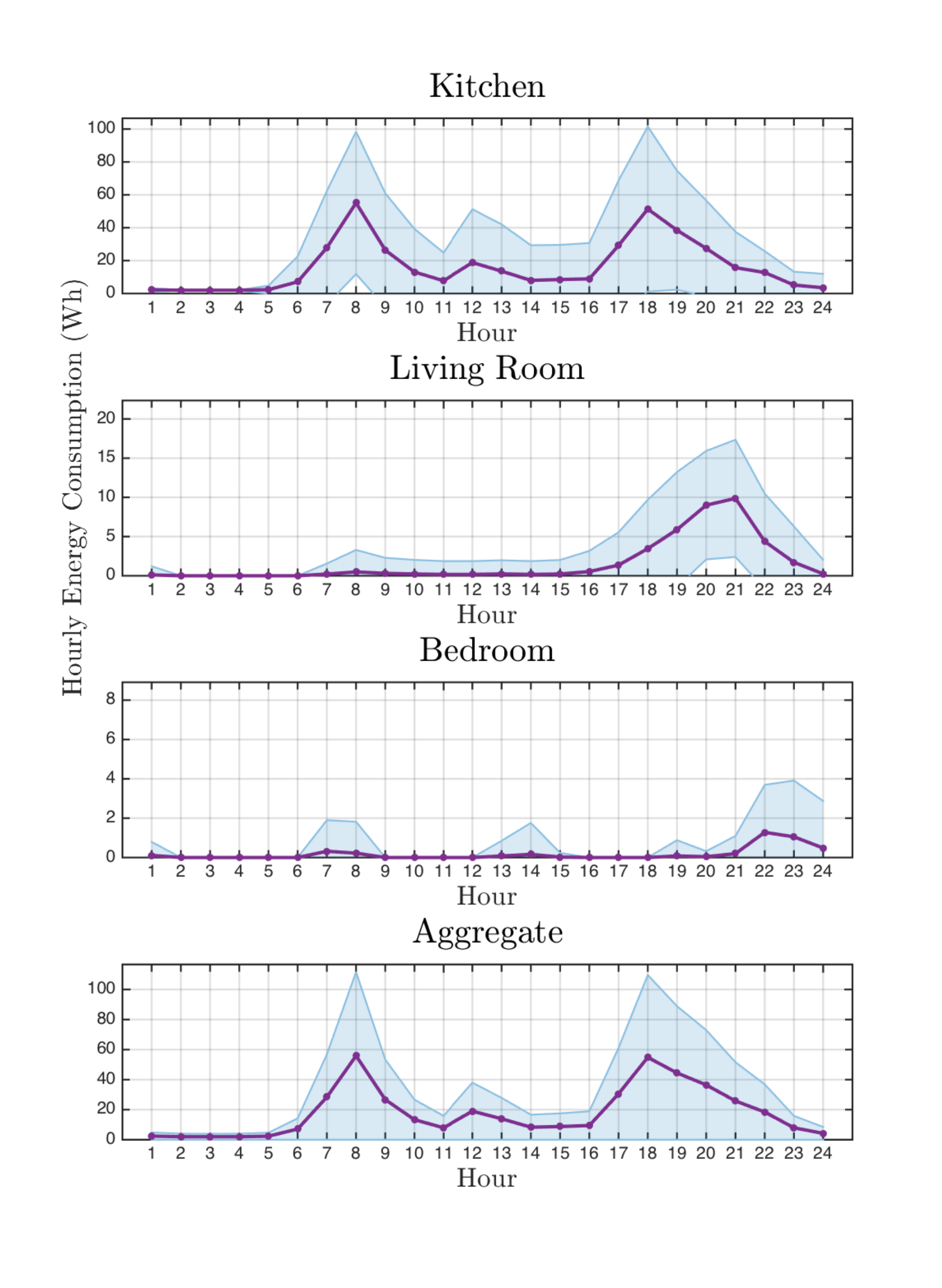}
\caption{Hourly usage patterns of main lighting appliances. Solid curve represents average consumption and shaded area represents one standard deviation above mean}
\label{fig:appProfiles}
\end{figure}

\subsubsection{Probabilistic Model Checking}\label{sec:pmc}

We experimented with soft constraints upper bounding the total power consumed during each hour.
Figure \ref{fig:appProfiles} depicts the hourly energy consumptions of each appliance, as well as the aggregated consumption, averaged across each day in the training data.
The maximum hourly consumptions occurring in the training data are not ideal bounds to use as soft constraints, since they tend to be trivially satisfied by the improviser.
Instead, for each hour $h$ we imposed a tighter bound $\Delta_{max}^h$ on the aggregate power consumption during that hour, where $\Delta_{max}^h$ was one standard deviation above the mean consumption in hour $h$ in the training data. 
Note that 89.2\% of the training samples were within this bound.  
The values $\Delta_{max}^h$ are plotted as the shaded curve at the bottom of Figure \ref{fig:appProfiles}.

% \begin{table}
% \begin{tabular}{ll|ll}
% \hline
% Hour of day (h) & $P_{max}^h (Wh)$ &Hour of day (h) & $P_{max}^h (Wh∫)$\\
% \hline\hline
% 1 &     4.50 & 13 &   42.73\\
% 2 &     2.25 & 14 &   29.82\\
% 3 &     2.00 &15 &   30.1\\
% 4 &     2.25 & 16 &   31.40\\
% 5 &     4.78 & 17 &   70.91\\
% 6 &    22.6& 18 &  106.68\\
% 7 &   63.27& 19 &   84.02\\
% 8 &   99.12& 20 &   67.77\\
% 9 &   61.86& 21 &   49.62\\
% 10 &   39.52& 22 &   34.216\\
% 11 &   25.02& 23 &   19.02\\
% 12 &   51.69& 24 &   13.20\\
% \hline
% \end{tabular}
% \caption{Hourly Energy Properties (one standard deviation above mean) Learned from Training Power Consumption Traces}\label{table:plimits}
% \end{table}

To compute the probability of satisfying these constraints,  we used the PRISM model checker~\cite{KNP11}.
As detailed in Section~\ref{encoding-prism}, the EDHMM and a monitor tracking hourly power consumption can be written as a discrete-time Markov chain.
This description translates more or less directly into the PRISM modeling language.
Having done this, the soft constraints can be put directly into PRISM using the PCTL formulation explained in Section~\ref{soft-constraints} to obtain the hourly satisfaction probabilities $1-\boldsymbol{\epsilon}_i,~i=1,\dots,24$.

% The model $W_t$
% % The two pieces of data-driven constraints, $\{ P_{max}^h\}$ and 

% Then, the learned EDHMM $\mathcal{W}_I$ and the power constraints $\{ P_{max}^h\}$  are then fed into a model checker. We use the tool {\sc PRISM} \cite{KNP11} to perform model checking given a PFA representation of the trained EDHMM and the PCTL formulas as defined in (\ref{eqn:pctl}). Complete properties for the training stage are given by Table~\ref{table:exp}. 

% For completion strategies $f_1(\cdot)-f_3(\cdot)$, the role of the PRISM model checker is to validate whether $\mathcal{W}_I$ is a valid control improviser, given $p_{min} \in [0,1]$. 
% For completion strategy $f_4(\cdot)$, PRISM is used for synthesizing the missing parameters of $A_I(h)$, assuming a sparsity pattern learned from the input trace, and illustrated in Figure \ref{fig:sparsity}.

% Data stream processing, EDHMM training and execution of the resultant control improviser is implemented using the Ptolemy II heterogeneous modeling and simulation environment \cite g{Ptolemaeus:14:SystemDesign}. Ptolemy II is also used to code generate a custom prism model that encodes $\mathcal{W}_T$, and produces the PCTL formulas to be checked on the model. The state machine in Figure \ref{fig:workflow} is taken from the Ptolemy implementation model and follows the modal model semantics in Ptolemy as presented in \cite{Ptolemaeus:14:SystemDesign}.
% \insertfigure{0.8\columnwidth}{ptolemy_modal}{EDHMM learning workflow}{fig:edhmm_workflow}
% \insertfigure{\textwidth}{workflow}{}{fig:workflow}

\subsubsection{Scenario-Based Model Calibration} 
As mentioned above, we tested three types of improvisers:

\begin{itemize} 
\item
{\bf Scenario I: Uncalibrated Improviser.}
This improviser uses the learned EDHMM with no model calibration.

\item {\bf Scenario II: Duration-Calibrated Improviser.}
This improviser uses the duration calibration heuristic described in Section~\ref{sec:heurstics}.
From the aggregate power profile given in Figure~\ref{fig:appProfiles}, we identified peak power consumption as occurring during hours 7, 8, 9, 17, 18, 19, 20, and 21.
For these hours, the probabilities of event durations greater than 60 minutes were set to zero and the distributions re-normalized.
Figure~\ref{exampleDurationDistributions} shows a sample set of original and calibrated event duration distributions for the 19$^{\text{th}}$ hour of the day.
%, i.e. between 7:00--7:59 pm.

\begin{figure}[!htb] \centering
\includegraphics[width=0.8\columnwidth]{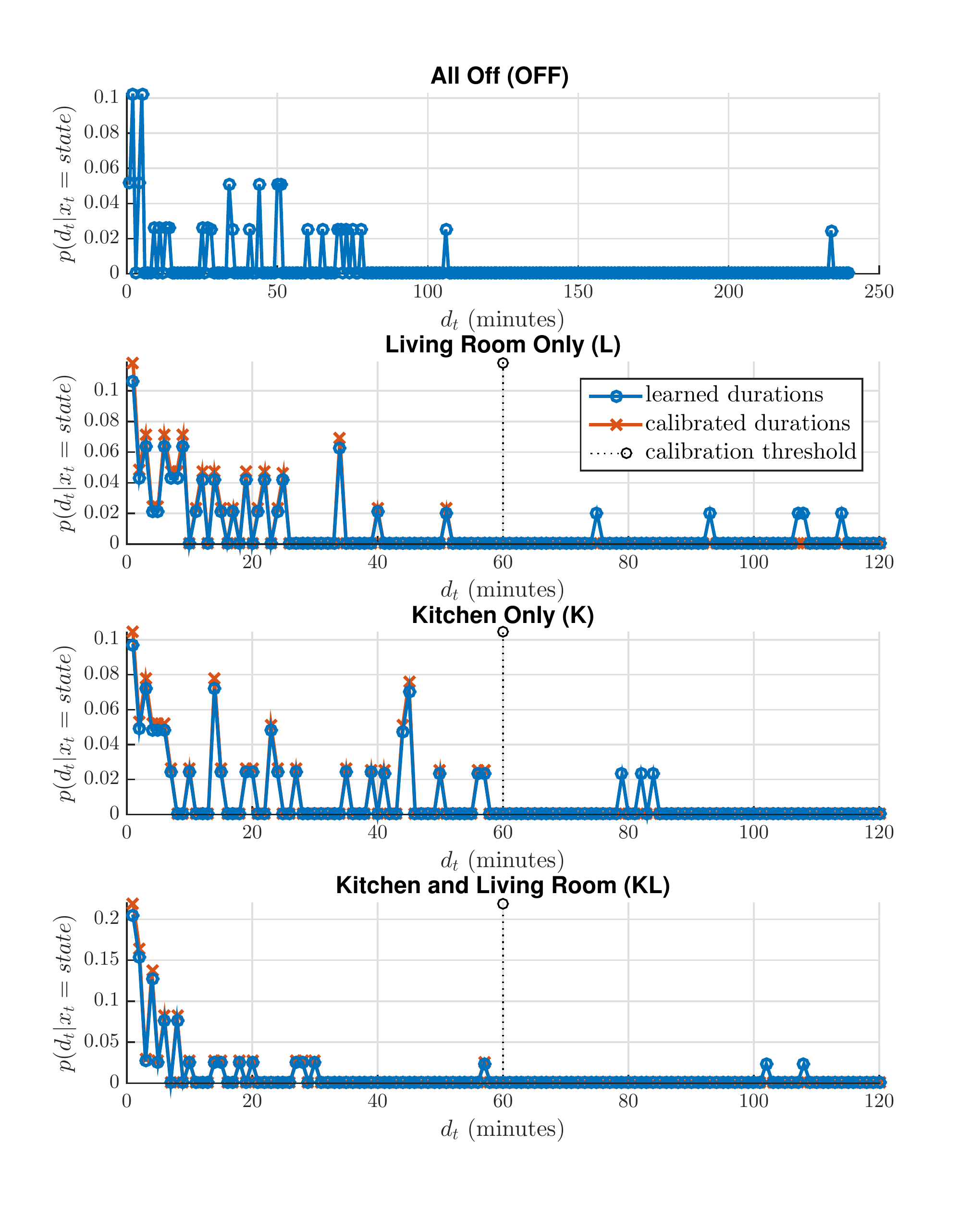}
\caption{Sample learned and calibrated duration distributions for h=19}
\label{exampleDurationDistributions}
\end{figure}

\begin{figure}[!b]\centering
\includegraphics[width=0.8\columnwidth]{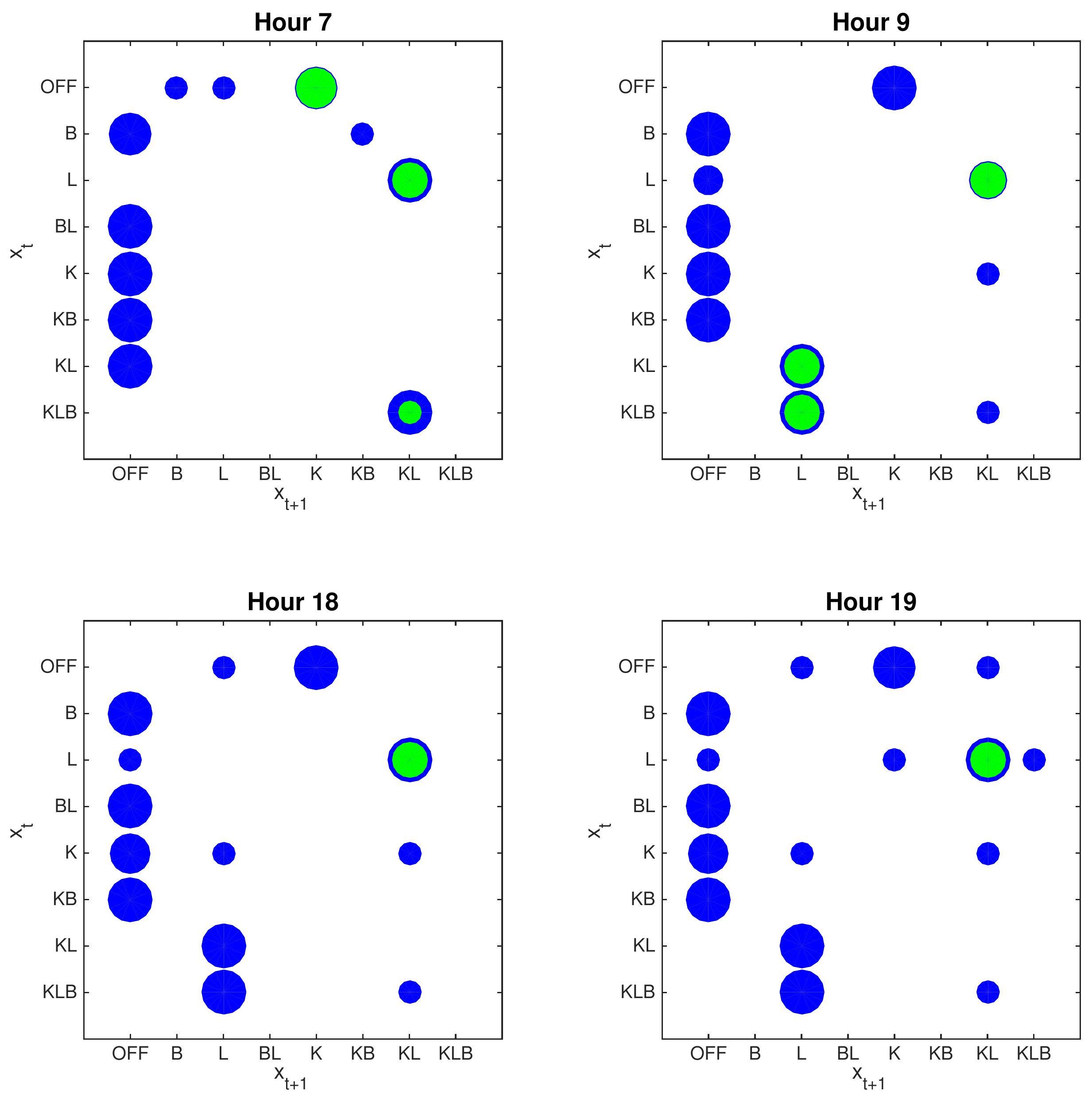}
\caption{Learned vs. calibrated state transition probabilities for selected hours. (Blue: Learned $A_h$, Green: Probabilities adjusted by Scenario III, See Table~\ref{table:exp} for label descriptions)}\label{fig:sparsity}
\end{figure}

\item
{\bf Scenario III: Transition-Calibrated Improviser.}
This improviser extends the previous one by also applying the transition calibration heuristic described in Section~\ref{sec:heurstics}. 
The set of hours for which transition probabilities were calibrated includes the peak hours considered in the previous section, with the addition of hours 4 and 5, for which very few events were recorded in the training data. As Figure~\ref{fig:appProfiles} indicates, the significant sources of power consumption are the kitchen and the living room lighting appliances. Therefore, we choose $x_r$ to include states K, L, KL, and KLB (see Table~\ref{table:exp} for label descriptions).

Figure~\ref{fig:sparsity} depicts some hourly transition probability matrices before and after calibration.
Each circle indicates a nonzero transition probability from state $x_t$ to $x_{t+1}$, where its area is proportional to the probability.
The blue circles show the original learned probabilities, and the green circles show the probabilities decreased by calibration.
For clarity, we do not show the corresponding increases in the probabilities of transitioning to the OFF state.
\end{itemize}
% \red{ILGE: check the comments here to see if you want them in III-F}
%We consider a scenario which essentially would yield a ``low-power mode improviser'' that produces traces similar to the training data, yet have provably more desirable power characteristics as ensured by the soft constraints. 
%The heuristic we adapt constrains transition probabilities to higher power consumption states to ensure the probability of violating the maximum power constraints is reduced.

\subsection{Experimental Results}

\begin{figure*}[!tb]
\centering
\includegraphics[width=0.9\textwidth, clip]{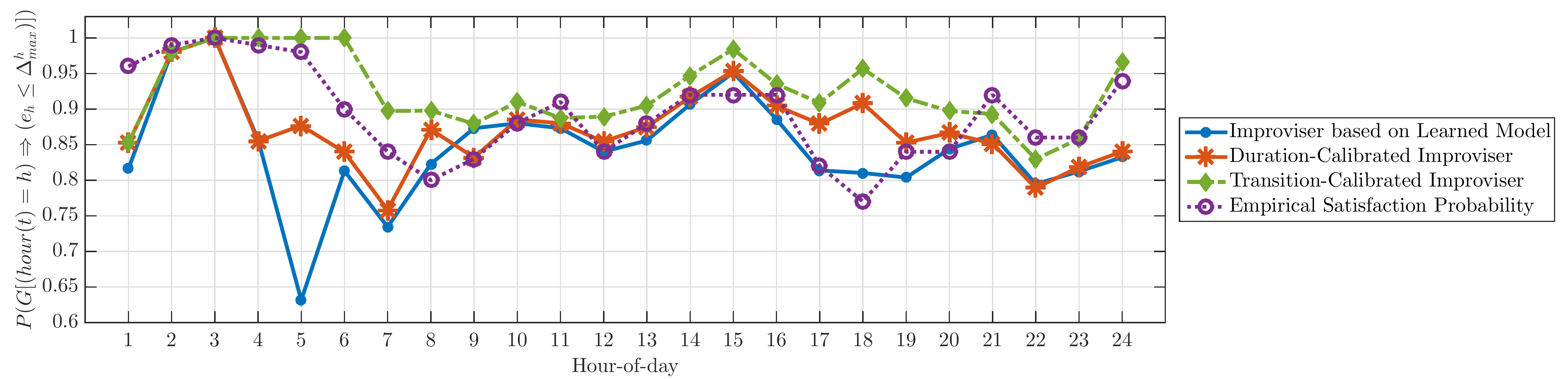}
\caption{Model checking results on the satisfaction probabilities of hourly soft constraints}
\label{pctl}
\end{figure*}

%%
%The goal of our experiments is to evaluate and compare the satisfaction probability of hourly power specifications (i.e., soft constraints) for the synthesized improvisers. 

%While, in general, the validity of an improviser with respect to soft constraints can be verified as explained in Section~\ref{sec:soft-constraints}, 

Our focus in this section is to evaluate the performance of synthesized improvisers using probabilistic model checking and to compare them based on their fidelity to soft constraints. It is also of interest to study the power profile characteristics of improvised traces to ensure scenario-based calibrations do not impact the similarity of improvisations to recorded data based on human actuation. 

%fidelity of the improvisers  satisfaction probabilities ($1-\epsilon_h$) are improved by applying scenario-based calibration strategies and to compare the performance of several improvisers based on their fidelity to soft constraints. 

Figure~\ref{pctl} summarizes model checking results for 
the original EDHMM and for the two calibrated models with constrained power consumption properties. We additionally provide the empirical probability of soft constraints being satisfied by training data, mainly for visual comparison. 
Model checking results suggest that the improviser based on the learned EDHMM behaves comparably to the empirical satisfaction probabilities, however, since the soft constraints are not explicitly enforced by the EDHMM, some hourly probabilities significantly deviate from empirical values. 
When we investigate model checking results for the two calibrated improvisers, which aim to improve the probability of satisfying soft constraints, 
%and to quantify the performance of our scenario-based calibration heuristics.
we observe that the transition-calibrated improviser yields highest satisfaction probabilities on the soft constraints 
for all hours of the day.
%, with a minimum hourly satisfaction probability of 0.8281.
%Model checking results for the learned improviser suggest that soft constraint satisfaction probabilities for some hours (e.g., h=5,7) are dramatically low, compared to the overall behavior of the improviser. We next perform probabilistic model checking on the two calibrated improvisers to quantify the improvement in the satisfaction probabilities. These results are also provided as part of Figure~\ref{pctl}. 
The duration-constrained improviser 
%exhibits higher satisfaction probabilities for 
performs better than the learned model, for all hours except for hours 9, 21 and 22.
As explained in Section~\ref{sec:heurstics}, the duration  heuristic does not guarantee an improvement in the probability of satisfying the soft constraints.
This can be explained in this particular case by the phenomenon that at these particular hours, the state transition matrix tends to make transitions to high-consumption states more probable, and skewing the duration distribution towards zero causes more state transitions to be made during peak hours. 
%Since trimming the duration distributions increases the probability of making more state transitions during peak hours, high-power states are visited more often.

\begin{figure}[!b]
\vspace{-0.2cm}
\includegraphics[width=0.95\columnwidth]{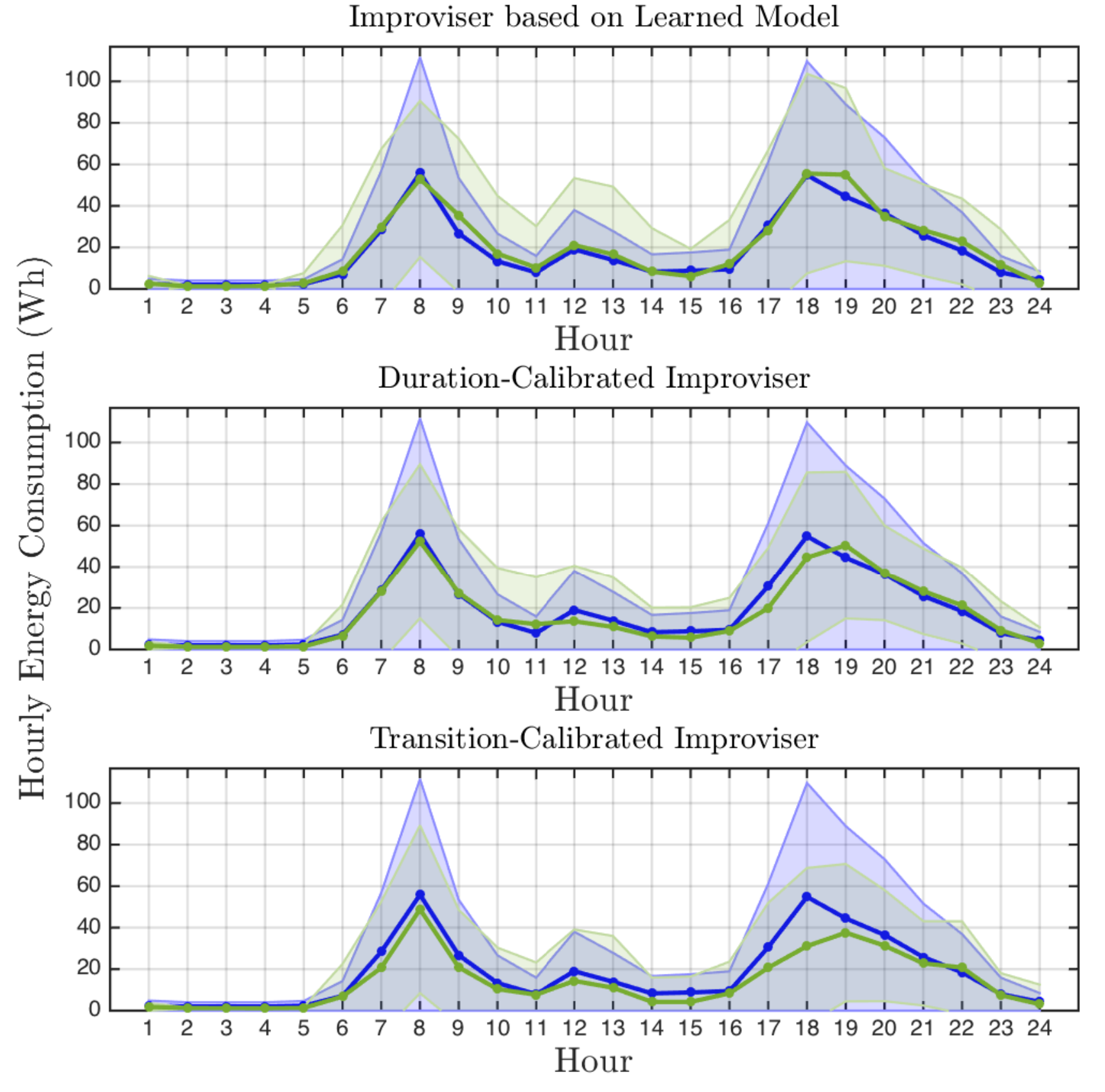}
 
\caption{Comparison of aggregate hourly energy profiles (Blue: Training data, Green: Improvisations. Solid curves represent mean energy, shaded region represents one standard deviation from mean)}\label{newHourlyProfiles}
\end{figure}

%Next, we are interested in comparing the hourly power profiles of traces generated by the resulting improvisers with those of training data.  
Figure~\ref{newHourlyProfiles} compares the aggregate hourly power consumption profiles obtained from the training data, with ones obtained from 100 20-day long improvisations generated by a particular lighting improviser. 
For all three improviser profiles, the hourly mean power trend matches that of the original data.
Moreover, for calibrated improvisers, the one standard deviation curve above mean mostly remains within the same bound for the original data.
%, showing that the calibrated improvisers also empirically achieve the goal of exhibiting desirable power properties.
%The transition-calibrated improviser is shown to produce traces that satisfy the one sigma power constraint more strictly than the duration-calibrated improviser, for each hour of operation.
%Specifically, e
Even though the duration-calibrated improviser has eliminated most of the highly variable power consumption trend exhibited by the uncalibrated improviser, it still demonstrates high variability in power for the hour range $h=9,\dots,12$ compared to the training data.
This behavior is successfully mitigated by the transition-calibrated improviser, which is shown to satisfy the one sigma power constraint more strictly than the duration-calibrated improviser as expected.

\begin{figure*}[tb]
%\begin{figure}[!ht]
\centering
\includegraphics[width = \textwidth]{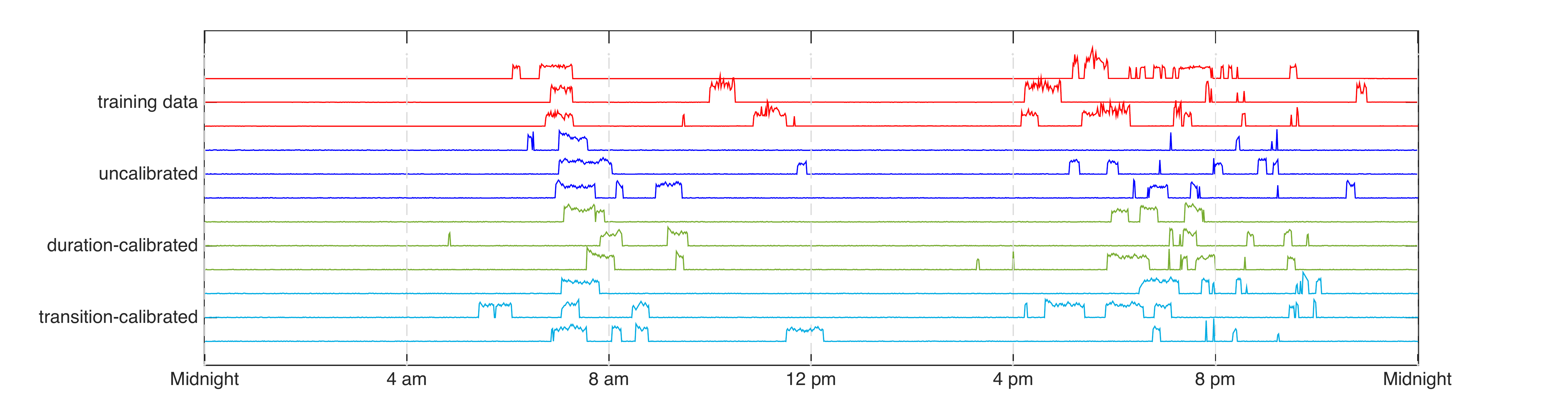}
\caption{Appliance 1 activation patterns over day-long intervals}
\label{fig:temporal}
%\end{figure}
\end{figure*}

Finally, in Figure \ref{fig:temporal}, we show several day-long traces from the three improvisers together with time-aligned excerpts from the training data.
Note that the uncalibrated improvisations are visually quite similar to the training data, illustrating the quality of the EDHMM as a model.
The calibrated improvisations are also qualitatively similar to the training data, but somewhat sparser as we would expect from enforcing constraints on power consumption. 
This demonstrates how our model calibration techniques are effective at enforcing soft constraints without drastically changing system behavior.

Overall, experimental results suggest that, given a suitable learning model, it is possible to synthesize a control improviser which produces randomized control sequences that are faithful to observed system behavior. More importantly, scenario-based model calibration methods can be applied to systematically constrain the nature of randomness, which is quantifiable via probabilistic model checking. Our experiments have shown significant improvements on the satisfaction probabilities of soft constraints after applying heuristic calibrations, while preserving desired qualitative characteristics in improvised control sequences. 

\section{Related Work}\label{sec:relwork}

{Control improvisation is an automata-theoretic problem that
was formally defined and analyzed in {\cite{fremont_et_al:LIPIcs:2015:5659}}. CI was applied to machine improvisation of music in {\cite{donze2014machine}}, where a symbolic reference melody was used to synthesize an automaton that was composed with a specification automaton (capturing user-specified musical properties) to produce a control improviser. In this work, we consider a case study in the field of home automation, which enables us to learn a more general Bayesian model. We represent training data by modeling temporal progression and the stochastic characteristics of underlying events given noisy sensor data. Moreover, as an extension of our previous work, we learn specifications from user-generated data directly, and perform scenario-based calibrations on the learned model to
enforce formal statistical properties. }

%Data-driven extensions of controller synthesis for modeling uncertainties due to human interaction has been a recent topic of interest. In \cite{feng2015controller}, controller synthesis of semi-autonomous UAV control with uncertain human operator behavior has been studied, and a possible extension of the framework with integration of data-driven operator models has been proposed. Similarly, in \cite{sadigh2014data}, a data-driven methodology to model and verify human driver behavior has been introduced, where data clustering based models of human modes of driving are built and verified for safety.  

Appliance modeling in residential settings has several proposed
benefits, including reduced power consumption, automated actuation of
smart appliances subject to energy pricing, microgrid load balancing,
and home security \cite{swan2009modeling}. Additionally, 
personalized advisory tools have gained popularity to provide adaptive demand-response prediction \cite{dr-advisor,li2012automated}.  
Bayesian modeling techniques for home appliance load modeling has been an emerging
topic of interest \cite{7049322}, and EDHMM-based models have previously been proposed for load disaggregation 
%for home appliance load modeling  
\cite{guo2015home}. Markov modeling of uncertainties in demand and energy pricing has been studied in \cite{neill2010residential}, which presents a reinforcement learning based approach to optimal load scheduling. 
%While our work shares the goal of quantifying uncertainty in human-actuated systems as above, our end goal is to guarantee formal properties on the resulting randomized controller.

%neill2010residential

%In \cite{sou2011scheduling}, a mixed integer linear programming based appliance scheduler is introduced, where user interaction is limited to a time window preference for each appliance to be active, and the control rule does not capture probabilistic variation in user preference. 

The related subjects of data-driven occupancy prediction \cite{aerts2014method} and user behavior modeling for energy demand predictions have also been studied in recent years. In \cite{wilke2013bottom}, a stochastic model to predict time-dependent user activity was presented, while in \cite{AksanliSGSC15}, a data-driven approach was adapted for learning residential power profiles based on user-specific factors. 
%In contrast to these studies, we focused on building appliance models that require no prior modeling of appliances or their users, but build solely on streaming sensor data from home appliances. To achieve this goal, 
%instead of building a context model for user behavior, 
%we considered a context-agnostic Bayesian model that captures time-dependent appliance duration and power consumption distributions, as well as time-dependent transition probabilities for appliance activation.
Integration of suitable occupancy and user prediction techniques with ours is a clear direction for future work.

%Our approach in this paper shares the vision of data-driven modeling of human-actuated systems, however, instead of synthesizing a Markov model of an underlying human behavior, we are interested in capturing the temporal progression of distributions associated with a semi-Markov pattern, as captured by the EDHMM. Our control objectives also vary, in that, we are interested in synthesizing a {\em randomized} controller that captures the probabilistic nature of a learned set of behaviors, as opposed to synthesizing optimal control traces based on hard specifications.
%

\section{Conclusion}
\label{sec:conclusion}

In this paper, we address the problem of randomized control for IoT
systems, with a particular focus on systems whose components
can be controlled either by humans or automatically.
From streams of time-stamped system events, we learn models
that are assumed to vary as a function of an underlying state
space governed by events with durations. 
We leverage the recently-proposed technique of control improvisation \cite{fremont_et_al:LIPIcs:2015:5659,fremont2014control}
%Using these, we show how to
%obtain improvisers that 
to generate randomized control sequences, which are {\em similar} to an
observed set of behaviors, and moreover, always satisfy some
desired {\em hard} constraints and {\em mostly} satisfy {\em soft}
constraints, while exhibiting {\em variability}.  
%A control
%improvisation problem on an explicit-duration hidden Markov model, and
%soft constraints that ensure  
%satisfaction of hourly probabilistic constraints on power
%consumption. 
We presented an implementation of the end-to-end
control improvisation workflow using the PRISM tool to enforce
soft constraints on the improviser.
We evaluated our technique in the domain of home appliance
control by synthesizing improvisers to control a group of lighting appliances based
on learned usage patterns and subject to probabilistic constraints on power consumption.
The results of our experiments showed that our methods can effectively enforce soft constraints while largely maintaining qualitative and quantitative properties of the original system's behavior. 

For future work, we plan to investigate new applications of this
framework in the IoT space. We also plan to investigate techniques to
improve the efficiency of our scheme, as well as its
implementation on real hardware.

\section*{Acknowledgment}
{This work is supported in part by the National Science Foundation Graduate Research Fellowship Program under Grant No. DGE-1106400 and by TerraSwarm, one of six centers of STARnet, a Semiconductor Research Corporation program sponsored by MARCO and DARPA.}

\bibliographystyle{IEEEtran}
\bibliography{root} 
 
\end{document}